# Complementary Theories of Energy Gaps in HTSC

J. C. Phillips[1*], J. Jung[2], D. Shimada[3], and N. Tsuda[4]

1. Bell Laboratories, Lucent Tech. (Retired), Murray Hill, N. J. 07974-0636, USA
2. Dept. Physics, Univ. of Alberta, Edmonton, AB  T6G 2J1, Canada
3. Dept. Physics, Shizuoka Univ., Shizuoka 422, Japan
4. Dept. Appl. Physics, Science Univ. Tokyo, Tokyo 162, Japan

We examine experiments on energy gaps in high temperature superconductors (HTSC) in terms of experimental probes that utilize momentum, position, or neither. Experiments on very high quality optimally doped BSCCO mechanical tunnel junctions show a sharp energy gap with a maximum anisotropy of ~ 10%, while ultrahigh precision ARPES experiments show 100% anisotropy (d-wave pairing). We resolve this conflict by showing that the latter result is an artifact caused by the momentum-projective nature of ARPES and glassy orthorhombic dopant correlations.  The latter appear to be a universal feature of the intermediate phase that is responsible for HTSC.  Apparent large inconsistencies between position-projective STM gap data and tunnel junction data on severely underdoped and optimally doped BSCCO, with and without Ni and Zn impurities, are also resolved.

## I.     INTRODUCTION

The causes of high-temperature superconductivity (HTSC) are still mysterious, although more than 50,000 experiments have studied this subject.  Theories have generally focused on only a few (often only one) kind of experiment.  Here we discuss many kinds of decisive experiments, and compare interpretations of these from two *complementary and apparently contradictory* viewpoints: **K** space and **R** space.  The **K** space theories are based on the effective medium approximation (EMA) and treat *crystal* momentum either as a good quantum number, or treat Bloch waves as good first approximations to quasi-particle states near the Fermi energy $E_F$.  The **R** space theories ignore the host crystal structure and focus on *glassy* nanostructures (such as nanodomains) that are disordered.  These disordered structures arise as a result of ferroelastic misfit, and the nanodomain dimensions are quite small (~ 3 nm), and possibly diffusion-limited.  For concreteness we focus on two specific theories, the very popular **K** space theory involving a strongly anisotropic energy gap (d wave pairing[1]) which vanishes in certain directions, and the almost unknown **R** space theory involving percolating filaments (centered on partially disordered dopants[2,3]).

The differences between these two theories are not merely a matter of representations chosen for basis functions. In the *continuum* **K** space theory the states near $E_F$ are metallic *positive* energy states describable by plane waves and partial wave expansions, with anomalous properties generated by unknown forces which somehow favor strong forward scattering in crystallographically indexed directions[4]. In the *discrete* **R** space theory the states near $E_F$ are *negative* energy states lying in a pseudogap and describable by wave packets whose centers follow filamentary paths through dopants and around nanodomain walls containing pseudogaps. Along a curvilinear path $\mathbf{R} = \mathbf{R}_i(t)$, the local filamentary momenta $\mathbf{K}_f(\mathbf{R}_i)$ are oriented parallel to the local tangent $d\mathbf{R}_i/dt$, which often changes direction. Moreover, at any composition only part of the sample need be superconductive, while other parts can be locally insulating or locally normal metals. In the **R** space theory the filamentary paths avoid the nanodomain wall network by *zigzaging* along (100) and (010) intraplanar segments, as well as between layers through resonant tunneling centers, as shown in Fig. 1. This curvilinear motion erases all the planar angular anisotropies that might occur in **K** space theories; in the absence of strain, the energy gap becomes *completely isotropic* (100% s-wave).

## II.     EVIDENCE FOR ENERGY GAP ANISOTOPY

The anisotropy of the energy gap $E_g = 2\Delta$, which is the central feature of the d-wave theories, can be established best by probing spectroscopically on nearly ideal (atomically flat, step-free) surfaces of BSCCO with two very different kinds of experiments. (In practice all high-resolution studies of $\Delta$ are carried out on BSCCO, because it is possible to prepare nearly perfect surfaces with this micaceous material.) In angle-resolved photoemission experiments (ARPES)[2] the probe is a high-energy (~ 22 eV) photon. The surface momentum $\mathbf{K}_s$ of the emitted electron (whose energy is lifetime-broadened) is measured; the best resolution obtained so far is ~ 5 meV. In conventional tunnel junctions[5] the voltage is that of $\Delta$ (~ 25 meV) or $\Delta$ plus a phonon energy $\omega$ ($\Delta + \omega$ ~ 75 meV), so that the energy of the probe is about *$10^3$ times smaller* in the junction experiments, which have much better resolution (~ 1meV). The tunneling characteristics $dI/dV$ and $d^2I/dV^2$ are averaged over $\mathbf{K}_s$ and $\mathbf{R}_s$, and so are unbiased towards either $\mathbf{K}_s$ or $\mathbf{R}_s$, as they depend only on eV = E($\alpha$) – $E_F$, and the quasiparticle states $\alpha$ in general are not indexed by either $\mathbf{K}_s$ or $\mathbf{R}_s$.

At first it might appear that the tunneling experiments, because of their much better resolution, would intrinsically yield much more informative results than the photoemission experiments. However, the tunneling experiments involve some kind of electrical contact with a metallic probe (either a metallic electrode or a microscope tip), and such contact can deform the soft outer layer of the micaceous BSCCO sample (*strain broadening*). This means that the results obtained by tunneling can vary not only with the quality of the sample cleavage, but also with the nature of the contact. In order to ascertain the extent of the uncertainties introduced by this variability, it is desirable to survey a wide range of data obtained with a variety of tunneling configurations, as we will do in this paper. Such a survey involves certain guiding principles. The most important of these is that because entire functions, I(V) and its first and second derivatives, are measured, rather than only isolated numbers, the internal consistency of the results obtained with a given contact configuration themselves exhibit the extent of strain broadening. In the best cases this strain broadening can be very small, yielding extremely valuable results. In other cases the strain effects are larger, but exhibit very interesting chemical trends with doping. These chemical trends reveal the nature of the electro-mechanical coupling of BSCCO sandwiches as a function of dopant density. We will systematically analyze these trends here; our analysis shows that, although the experiments are difficult and must be carried out with great care, the results obtained to date are far more informative than has often been believed to be the case. In general, because of their higher resolution, tunneling experiments have provided the best information obtained to date on electronic structure near the energy gap in HTSC. Moreover, the contact problems themselves provide additional insight into the nature of electromechanical interlayer coupling.

The photoemission (ARPES) experiments indicate an anisotropic $\Delta$ in the ab plane which ranges from 0 [110] to 20(5) [100] meV, as expected for d-wave anisotropy. The mechanical junction experiments on the *best BSCCO junction* indicate an extremely sharp $\Delta$, as shown in Fig. 2. This gap exhibits almost no anisotropy (an upper bound of 3 meV, which itself may be associated with residual damage or strain) for an s wave gap of 24.5 meV.

### III.    HIDDEN SUPERCONDUCTIVE FILAMENTARY STRUCTURE

In d wave theory these contradictions might somehow be resolved by assuming that the junction surface conductivity is extremely anisotropic (several orders of magnitude), completely unlike the normal state bulk conductivity, as a result perhaps of the usual unknown interactions.  In the filamentary model, because **K** is not a good quantum number, the angular variations in photoemissive yield *are not intrinsic signatures* of gap anisotropy, and instead can only arise from another *extrinsic* cause.  Here we identify this hidden mechanism as dopant self-organization relative to a nanodomain matrix.  In order best to observe dopant self-organization, one needs large single crystal samples, which are not available for BSCCO.  Such samples exist for superconductive ($Ba_{0.6}K_{0.4}BiO_3$), as shown[5] in Fig. 3, and ferromagnetic ($La_{1-x}Sr_x$ MnO) doped perovskites, as well as in layered superconductive pseudoperovskite cuprates $La_{2-x}Sr_xCuO_4$ (x = 0.15) and $YBa_2Cu_3O_{6+x}$ (x = 0.20, 0.35, 0.60, and 0.93).  In all these cases anomalous features (dispersive gaps and pseudogaps) have been observed by inelastic neutron scattering in the [100] LO (longitudinal optic phonon) bond-stretching vibrational spectra[6].  (No anomalies are observed along [110] directions.  Also the anomalies correlate closely with the superconductive regions of the phase diagram, and are absent in the normal regions, as shown in Fig. 3.)

These [100] LO phonon anomalies are explained by a microscopic theoretical model which is compatible with the only known satisfactory fundamental model of superconductivity, the BCS theory.  The model assumes that the dopants are orthorhombically self-organized into local (100)-oriented chains and [100] planes parallel to Cu-O bonds (clusters only, that is, static medium-, not long-, range order)[7].  Local (glassy) orthorhombicity, as measured by XAFS, is associated with (100) self-organization.  In $La_{2-x}Sr_xCuO_4$ for some x there is local (or short-range) orthorhombic order which is larger than the macroscopic orthorhombic distortion of the unit cell[8], and *only when both are present and different*, that is, 0.15 < x < 0.21, does one have bulk

HTSC. In this case the flexibility provided by orthorhombically *disordered* nanodomains is clearly a key factor that makes HTSC possible.

With this information in hand, one can easily explain the otherwise mysterious anisotropic gap observed by photoemission[1]. Parallel to dopant chain segments the full gap $\Delta \sim 20$ meV is observed[1]; in this geometry $\mathbf{K_s}$ is parallel to [100] along the chain segment. From the sharpness (Fig. 3) of the pseudogaps in the [100] LO spectra[6,7], one can infer that these chain segments are at least $d \sim 6a \sim 3$ nm long, where $a \sim 0.5$ nm is the unit cell dimension. As we shall see below, this is just the nanodomain diameter. According to the geometry for *zigzag* paths shown in Fig. 1, this is usually the maximum filamentary segmental ab planar *zig* length possible before the path *zags* parallel to the c axis through a resonant tunneling center to a nanodomain in an adjacent metallic plane. Along such a segment, similar to a chain segment in YBCO, the oscillator strengths for excitation of photoelectrons can be enhanced by coherence of the amplitudes along the chains, an *antenna effect* that is taken for granted to hold for **all** excitations of any orientation in theories where $\mathbf{K}$ is *assumed* to be a good quantum number. Because of this selective antenna effect, for $\mathbf{K}$ along [100] and initial electronic energies near $E_F$, excitations from the filamentary states will dominate all others. The apparent $\mathbf{K_s}$ - biased ARPES value of the gap parameter $\Delta_p$ should be about the same as that measured for the unbiased tunnel junction $\Delta_j$. Indeed there is good agreement between these two values ($\Delta_p([100]) = 20(5)$ meV and $\Delta_j = 24.5 (1.5)$ meV).

What happens when $\mathbf{K_s}$ is parallel to [110]? In this case, the overlap between the initial filamentary antenna state, with local momentum parallel to [100] or [010], and the excited [110] final state, is greatly reduced, and the antenna enhancement of the oscillator strength is reduced. There are two contributions to this reduction. First, because the final state energy is 22 eV ~ 20W, there can be as many as 20 oscillations of the final state wave function in one unit cell. However, many filaments may extend across a nanodomain of 6 unit cells. Because of this second factor the oscillator strength may be enhanced by a factor of 100 parallel to a filament compared to transverse.

An alternative process that would contribute to the measured spectrum is excitation of carriers from overdoped regions that are Fermi liquids with zero gap. The volume of such regions is less than that of the superconductive regions, and the oscillator strength is smaller too, as the currents are less coherent. Such regions produce a featureless current that would be treated as background in analyzing photoemission data for $K_s$ in the [100] orientation, where the antenna effect is strong. However, for $K_s$ in the [110] orientation, where it is much weaker, as the coherent filamentary currents are much weaker, the contributions from Fermi liquid regions appear as an effective gap of zero energy. It is understandable that the possibility of this anisotropic enhancement of the oscillator strengths was overlooked when the photoemission data were first observed, as at that time many of the effects associated with (100) orthorhombic dopant ordering had not been reported. However, now the data base supporting (100) dopant ordering is quite substantial[2,3], and the possibility that the ARPES data were overinterpreted must be given serious consideration. Put differently, we now have good reason to believe that the large tetragonal anisotropy thought to exhibit a d-wave energy gap was actually a large anisotopy in the oscillator strength for excitation of 22 eV electrons.

Anisotropic structure, ascribed to d waves, has also been observed in Andreev scattering experiments on YBCO grain boundary junctions fabricated by pulsed laser ablation[1]. The absence of the field and temperature dependence predicted by $K$ space theory from the observations has, however, raised doubts[9]. Perhaps the observed oscillations are associated with extrinsic static nanostructure of the interface.

## IV. TUNNELING SPECTROSCOPY

The question of whether to use the $K$ space or the $R$ space model has also appeared in conjunction with scanning tunneling microscope data, which seem to be ideally suited to the $R$ space approach. Early data[10,11] have been interpreted as supporting the d wave $K$ space description, but recently a different note has been heard, with much in common with the non-translationally invariant, dopant-centered $R$ space approach[12]. These experiments observe the centers and angular variation of states pinned to a narrow energy region near the Fermi energy. There are two cases of such nanoscale zero-bias defect or impurity states: native X defects[10], probably associated with O vacancies spaced 1.4 nm apart[12], which can act as dopants supporting HTSC, and extrinsic Zn impurities[11] which quench HTSC. According to the d-wave $K$ space model, the radial extent of the nearly zero-bias states should depend on the nature or strength of the impurity (X or Zn) potential. On the other hand, the angular behavior should be characteristic of the host d-wave pairing itself and should be *independent* of the nature of the impurity. The nanodomain **R** space model predicts just the opposite: the radial extent of the zero-bias states depends only on the host composition-independent[10], ferroelastically fixed nanodomain size of 3.2 nm, which is in good agreement with earlier estimates[13,14] of nanodomain dimensions in YBCO. Contrary to the **K** space model, this measure of the radial extent of the nearly zero-bias states should be nearly independent of the nature of the impurity, as observed for X and Zn.

The nanodomain **R** space model predicts that the angular behavior is determined by whether or not the impurity is located within the domain, or at a domain wall with 2-fold symmetries, or wall intersection with 4-fold symmetry.. The X impurities are native and are most likely to be found in the domains. The natural site for the misfit Zn in the $CuO_2$ planes is segregated at the 4-fold symmetric corners of the nanodomains, as the nanodomain walls function as relievers of interplanar $CuO_2$-BaO ferroelastic misfit. Generally the bonding distortions at the domain walls must already be large, and so the strain energy will be minimized if the Zn occupies a more distorted wall site rather than an undistorted domain interior site. Moreover, if the Zn does even better, and occupies a corner site, because of the 4-fold symmetry it is likely to quench the local pseudogap associated with Jahn-Teller relaxation of the neighborhood of that site. This will convert the four associated nanodomains from coherent, negative-Fermi energy, filamentary tunneling regions, to positive-Fermi energy, Fermi liquid (effectvely overdoped) non-superconductive regions

Note that there are $\sim n^2$ unit cells/nanodomain, and that a corner Zn will affect four nanodomains. With nanodomain diameters of 3 nm and the planar lattice constant a ~ 0.5 nm, there are n ~ 6 unit cells in a nanodomain diameter. This means that ~ $1/(4n^2)$ ~ 1% Zn can quench both superconductivity and filamentary vibronic anomalies, in agreement with critical Zn concentrations as measured in both STM[11] and infrared and neutron experiments[15]. Of course, **K** space d wave pairing makes no quantitative predictions at all concerning the critical Zn concentration, because it is a descriptive continuum theory and contains no lattice scale for discussing chemical trends.

The Zn data[11] are especially noteworthy for the observed angular variation, which is much too rapid to be described by a merely quadratic d function, such as $F(x,y) = (x^2 - y^2)$, and a pseudo-gap based exponential would appear to give a much better fit. In addition to the zero-bias bright spot associated with each Zn atom, one can discern narrow tetragonal spokes emanating up to 3 nm from a spot. These spokes are not observed[10] in connection with impurity or defect X states that produce zero-bias bright spots in undoped BSCCO. In the **K** space model these spokes are regarded as evidence for d-wave superconductivity, which is supposed to be a characteristic feature of the host, and so should be observed in connection with *both* Zn states *and* X states. With an **R** space model these spokes represent both the anisotropy of the pseudogap which is localized[16] in the nanodomain walls, as well as the anisotropy of the Zn resonant charge density that is centered at the intersections of these walls. Then the large difference in anisotropy between the STM images for the X states and the Zn states is understood by the Zn site preference for nanodomain corners.

Localization of the pseudogap in nanodomain walls suggests the need for a mathematical model of such effects. Such a model is discussed in the Appendix, where it is used to analyze data on Ni impurities which appear to have segregated in the wall with 2-fold symmetries. Unlike the Zn case discussed above, the Ni impurities have a much less drastic effect on the pseudogap in their neighborhood.

Very similar STM data have been presented differently, and have been interpreted with a combined **K** space and **R** space model[16]. Pseudogaps have been identified as associated with narrow nanodomain ridges in **R** space. The interpretation of these data also utilized the **K** space d wave model to discuss gap broadening, but it seems to us that the gap broadening most probably reflects the strain produced by the interaction of the tunneling tip with the surface layer. The superconductive gap associated with domain centers is $\Delta_S$ = 42(2) meV. This value is nearly twice that of $\Delta_j$ observed in the tunnel junction and ARPES experiments. This is consistent with our picture, which says that isolated values of $\Delta_s(\mathbf{R})$ or $\Delta_s(\mathbf{K})$ are in general different from the average value of $\Delta_s$ which is measured in the mechanical junction tunneling experiment. [It has been shown[5] that analysis of the electron-phonon fine structure of the latter's $d^2I/dV^2$ by the Eliashberg equations gives $T_c$ = 86K. This value is in excellent agreement with the resistive $T_c$ = 88K. Moreover, the *entire function* associated with the phonon density of states is very well resolved, and is in excellent agreement with independent measurements by neutron spectroscopy, and theoretical calculations from a lattice dynamical model[5]. *These data leave no doubt at all that HTSC is caused by electron-phonon interactions, and that the average gap is sharp and isotropic*.] Thus the average value of $\Delta_s$ which is measured in the mechanical junction tunneling experiment at optimal doping is the standard or benchmark, and special considerations are required to explain anomalous values of $\Delta_s(\mathbf{R})$ or $\Delta_s(\mathbf{K})$.

Many tunneling experiments study $\Delta_s(\mathbf{R})$ in the surface molecular sandwich, which need not be the same as the bulk $\Delta$. This surface sandwich interacts with substrate sandwiches only through weak Van der Waals forces. It is this weak interaction that explains the easily cleaved, micaceous nature of BSCCO. In the surface BiO layer the boundary conditions for ferroelastic reconstruction are different from the bulk, where nanodomains form. In the surface sandwich buckling can occur, with pile-up of nanodomain walls to form widely spaced ridges, with widths of order the nanodomain diameter. (This pile-up is analogous to the formation of widely spaced bundles of steps (called risers) on slightly misoriented (viscinal) Si surfaces[17]. It can leave behind step-free, or wall-free, areas.) As a result of this buckling, the surface BiO layer can be electronically decoupled from the substrate layers, which include the CuO layer, where the electron-phonon coupling is much weaker ($\Delta_s(CuO) << \Delta_s(BiO)$) than in the BiO layer. Thus it is possible for the surface BiO layer alone to have $\Delta_S = 42(2)$ meV $>> \Delta_j = 24.5$ meV, as the latter is the filamentary average over both BiO and CuO layers.

So far we have discussed only tunneling spectra obtained on specially prepared Bi2212/pGaAs junctions with an epoxy-coated mechanical interface. Other ways of preparing Bi2212 tunnel junctions involve the "break" method, where the crystal is fractured and then the two pieces are mechanically rejoined, the point contact method, and Au electrodes. These different ways are illustrated in Fig. 4. In general these other methods have yielded some very interesting results, which we discuss below. However, they always produce gaps $\Delta$ which are broadened too much by strain to exhibit phonon fine structure in $d^2I/dV^2$ at $eV = \Delta + \omega$, as is the case with the epoxy-coated junctions. We believe that the epoxy coating of the Bi2212/GaAs junction relieves most of the stress associated with the free Bi2212 surface, and thus makes possible a nearly strain-free contact and a sharp gap. The epoxy acts as a strain-avoiding mechanical buffer, as shown in Fig. 4(c), in contrast to the direct contacts (including STM, which is also a point contact), which inevitably strain the surface, Fig. 4(b).

## V. UNDERDOPED MYSTERIES

The ($\Delta$, $T_c$) phase diagram probed by many Bi2212 break junctions and point contacts[18,19] is sketched in Fig. 5. It has puzzled many observers. The problem is that above optimal doping $x_0$, both $\Delta$ and $T_c$ go down, as expected from theory, but in the underdoped sample, while $T_c$ reverses direction and decreases, $\Delta$ continues to increase nearly linearly. Thus $T_c(x - x_0)$ has even parity, while $\Delta(x - x_0)$ has odd parity. In continuum theories (BCS) $\Delta$ and $T_c$ increase together monotonically, and in the strong-coupling limit are proportional to $\lambda$ and $\lambda^{1/2}$, respectively, where $\lambda$ is the electron-phonon coupling strength. Thus this "broken functional symmetry" is mysterious, and requires special consideration.

One possible resolution to this puzzle is the idea, mentioned above, that the upper BiO layer, or perhaps an entire formula unit sandwich, can be electronically decoupled from the $CuO_2$ layer or sandwich below it when the sample is underdoped. If the dopants support interlayer or intersandwich filaments, then when there are too few of these filaments, the intrinsic misfit between the surface layer, with its dangling surface bonds, and the lower layers, may lead to such decoupling. This kind of decoupling is also a kind of micaceous effect, but on an atomic scale. Similar effects occur at the free surfaces of semiconductors such as Si, and are responsible for surface reconstruction.

Another way of understanding the gap enhancement and $T_c$ suppression with decoupling is the following. When the outer layer or sandwich is decoupled from the substrate, the Cooper pairs can bind together more tightly because the amplitude of the phonon vibrations can increase, as they are no longer hindered by the substrate. That amplitude increase as the layer becomes more two-dimensional is well known in other contexts, as it leads to surface softening and pre-melting. When the outer layer is decoupled, it develops two surfaces, rather than one, so that it becomes even softer against vibrations normal to the surface. The enhanced vibrational amplitudes increase the electron-phonon coupling, which increases the gap at $T = 0$.

Is there any independent evidence for such decoupling in the data? As shown in Fig. 6, there is, in the point-contact phase diagram. For optimally doped and overdoped samples $\Delta(T)$ follows the BCS mean-field behavior. For the underdoped samples $\Delta(T)$ is too flat at low T, and then collapses[19] abruptly for $T/T_c$ above 0.8. This is suggestive of quasi-two dimensional behavior due to Kosterlitz-Thouless fluctuations[20], such as would occur when surface layer decoupling is present. This phenomenon has been studied in [xy model] spin systems in terms of a crossover from three- to two-dimensional magnetism. Collapse of the order parameter similar to that shown in Fig. 6 has been observed[21] in antiferromagnetic, graphite-diluted $CoCl_2$, and decoupling effects have been discussed extensively[22] for type II phases in a magnetic field.

Even more complex behavior of $\Delta(T)$ is observed in planar tunnel junctions utilizing Au electrodes[23]. Broadly speaking, there is more evidence for even stronger, possibly thermally activated, decoupling of the outer BiO layer for $T/T_c > 0.5$. This seems natural, because the interfacial Au-BiO bond should be much stronger than the micaceous BiO-BiO bond. As the interlayer filamentary dopant bonds weaken for larger $T/T_c$, Au contacts can lead to lift-off at the planar molecular level.

## VI. INFRARED AND RAMAN SPECTRA

Further evidence supporting the interlayer *zigzag* **R** space filamentary model is provided by the broken symmetry that has been observed in YBCO infrared vibronic spectra[24]. It was recognized[25] early that the changes in the vibrational spectra of $YBa_2Cu_{3-y}Zn_yO_{6+x}$ between $x = 0$ and 1, and $y = 0$ and 0.03, measured by neutron spectroscopy, were too large to be explained by conventional **K**-conserving force-constant vibrational (shell) models. Especially anomalous are the peaks near 40 and 50 meV. Their large chemical shifts in intensities can be understood as the result of vibronic correlations that arise when carriers and phonons are bound to an interlayer filamentary path[15]. These correlations are particularly large because when the carriers and phonons move parallel to the c axis, they are funneled through resonant tunneling centers in semiconductive layers where the electron-LO phonon interaction may be enhanced[7] by a factor as large as $n^2 \sim 30$.

Raman scattering experiments[26,27] have also been interpreted as showing evidence for d-wave gap pairing. Generally speaking, the selection rules for Raman scattering are weaker than those for infrared scattering, which means that the kinds of broken symmetry effects that are evident in the infrared experiments are much less obvious in the Raman spectra. The latter do, however, contain two additional features of interest. First, strong magnon-magnon resonant scattering bands are observed, in addition to particle-particle resonant scattering bands, peaking near 3 eV. The former extend well into the intermediate (HTSC) phase region, in contrast to antiferromagnetism (AF) itself, which is confined to strongly underdoped samples. In the **K** space continuum models, it is very difficult to reconcile these strong AF "dynamical fluctuations" (as they are described in continuum models) with HTSC, as described in many papers, for example, refs. 13-27, and also others, of ref. 27 here. Instead, if one simply abandons **K** space continuum models, and replaces them with the **R** space models that are the natural consequence of the ferroelasticity that has always been so characteristic of perovskites, there are no problems. The magnons are automatically localized in the domain walls, where there is a gap ~ 3 eV in the magnetic excitation spectrum associated with spin density waves. The second interesting feature of the Raman data, the observation of d-like $(x^2 - y^2)$ anisotropy in the pseudogap, is explained in the same way that we have explained the ARPES data, as the result of short-range *static dopant or domain wall* ordering with (100) or (110) orthorhombic symmetry.

## VII. INTERMEDIATE PHASE

The ultimate test for the validity of the **K** space or **R** space models lies in the phase diagram, which must contain a phase intermediate between the low-dopant density insulating phase and the high-dopant density Fermi liquid phase, that supports HTSC. It may seem easy to find such a phase, but if it is metallic and gapless, why is it not a Fermi liquid? Indeed, in the continuum EMA that is the basis for all **K** space models, there is only one phase transition, and no intermediate "strange metal" phase, hence no HTSC.

However, in **R** space models with dopants, it is easy to find such a phase, as the filaments that percolatively tunnel (at Fermi energies within the pseudogap in the EMA) form just such a phase[7,28]. Moreover, the phase transitions that define this phase have universal topological properties. Unlike the usual descriptive parabolic model for $T_c(x)$, the two transitions are specified, and are not symmetric. The low-density transition is always *continuous*, and corresponds to a classical percolative transition. The high-density transition occurs when the transverse filamentary localization energy becomes too large, so that self-organized[29] filaments no longer have lower free energy than the Fermi liquid; this transition is always *first order*. If the sample is not fully equilibrated, it may be difficult to recognize that this transition is first order. However, the first-order character appears to have been identified unambiguously[30] for YBCO and HgBCO[3,31].

## VIII. CONCLUSIONS

In conclusion, we have compared and contrasted two *complementary and apparently contradictory* theoretical platforms for HTSC, based on crystalline **K** space and glassy **R** space. Although the **K** space platform has been by far the more popular, we have identified many experiments of very high quality that cannot be explained consistently by the crystalline **K** space platform, but that are explained very well by the glassy **R** space platform. These include recent STM experiments, that would seem to be most suitable for the glassy **R** space approach, but which were initially explained by the **K** space d wave model. This means that consistent theoretical interpretation of all (50,000!) experiments in these complex materials is becoming a science of its own. It will be important in the future to consider not only the (hitherto) very popular **K** space models, but also the (so far) much less popular, but increasingly more relevant, **R** space models. It is also important to bear in mind that no experiment can prove a theory correct, but it can falsify a theory. There are many failures of continuum d-wave models that prove them incorrect, and none for the discrete filamentary nanodomain model.

*Postscript.* The validity of many of the assumptions of the filamentary **R** space model has been demonstrated elegantly[32] in an analogous context, that of semiconductor nanostructures based on a δ-doped GaAs/AlGaAs heterostructure. Current flow near a quantum point contact embedded in the two-dimensional electron gas is imaged by a scanning probe microscope. The observed current flow is both filamentary and coherent. The quantum point contact is operated in the lowest quantum step mode, and it plays the same role that the resonant tunneling centers do in Fig. 1. From this analogy one can conclude that normal-state filamentary transport in HTSC retains its coherent character in the presence of glassy disorder. This coherent filamentary character accounts for many normal-state transport anomalies[28]. The entire situation is summarized for the reader's convenience in a compact form in Fig. 7.

After this paper was completed, a different interpretation of the ARPES anomalies appeared[33], based on "electron fractionalization", as in the quantum Hall effect. We submit that our explanation, based on modification of oscillator strengths and lifetimes caused by short-range dopant ordering, although much less glamorous, is much more credible.

## APPENDIX

In Sec. IV a careful analysis of radial/angular variations of states near $E_F$ for native O vacancy and Zn substitutional impurities on Cu sites in BSCCO showed that the STM data could not be explained by a d wave model. Instead the observations were well fitted with an s wave gap by assuming that the Zn was precipitated on sites of intersecting nanodomain walls, where the Zn quenched the pseudogap associated with those walls. Here we turn to a different case, Ni substitutional impurities on Cu sites, also studied[34] by STM. Again the data are interpreted using a d wave model, and again it turns out that this continuum model contains serious internal contradictions that are resolved by combining an s-wave gap with the internal structure of nanodomains and filaments.

The central feature of the selected Ni data is that, quite unlike the Zn data, it still contains a gap of 28 meV, which is close to the average superconductive gap of 24 meV near optimum doping. Detailed data are shown of the spectral state density $d^2I/dV^2$ averaged over 16A squares centered on Ni and its two nearest Cu neighbor rings. These data exhibit two donor-like impurity states within the gap, and two acceptor-like states with the same absolute energies relative to midgap. This electron-hole symmetry is rather complex: the acceptor-like state densities are larger on the central Ni and the $2^{nd}$ Cu neighbors, while the donor-like state densities are larger on the nearest Cu neighbors. Apart from the impurity bound states, there is no evidence of d-wave gap filling, and the observations suggest an s-wave gap.

First one can note that orbitally Ni is much more like Cu than Zn is. Thus both CuO and NiO have octahedrally coordinated cations, while in ZnO the cation is tetrahedrally coordinated. Therefore one would expect that Zn could quench the wall Jahn-Teller pseudogap and could precipitate at the more stable 4-fold nanodomain corner. Ni could precipitate at a site of intermediate character in a nanodomain wall, and produce impurity states in the wall pseudogap, but with only 2-fold mirror symmetries Jahn-Teller distortions may still be present and the pseudogap may not be quenched.

It is not possible, in general, to solve exactly the problem of the nature of the zigzag filamentary states in the self-organized dopant impurity band (previously called the Stormer band[7]). However, one knows that because of their zigzag filamentary character **K** is not a good quantum number for these states. On the other hand, to carry out explicit quantitative calculations of impurity states one is almost forced to adopt oversimplified continuum models in which the impurity is embedded in a Fermi liquid[35] and **K** can be used to determine the angular and radial wave function around the impurity. If the band structure is two-dimensional and electron-hole symmetric, a logarithmic singularity in the density of band states exists at half-filling. If the impurity is represented by an on-site potential U and an exchange interaction J, one can then solve the scattering (T matrix) equations to exhibit the qualitative nature of the impurity states. With an s-wave gap and U ~ W, the impurity state energies are distributed over a wide energy range ~ W, whereas with a d wave gap they are localized within the gap. As the observed impurity energies near optimal doping generally lie in the gap, this seems to be strong evidence in favor of a d wave gap, *providing that the continuum hypothesis is valid and that plane-wave states indexed by* **K** *correctly describe the scattering properties of states in the effective medium.*

On the other hand, in the filamentary nanodomain model, all electrically active states near $E_F$ are reordered, and are associated with *discrete* dopant-centered filaments, with the energy scale being set by the width of the dopant impurity band, which is of order the pseudogap $E_p$ and/or the superconductive gap $\Delta$. This *automatically guarantees* that the impurity states will generally fall in the gap, and it requires an s-wave gap, as with a d-wave gap all impurity interaction energies would be further reduced by a factor at least of order $E_p/W \sim 1/20$.

The topographical map[36], Fig. 2(c), shows that the Ni atom is situated at the center of a largely dark, semiconductive valley parallel to the a axis, and that in this valley the pseudogap is modulated with a period of about 10-15A. (This explains the lack of full tetragonal symmetry in Fig. 2(b)). This dark valley could well represent the reconstructed surface image of a sub-surface stress-relieving semiconductive nanodomain wall. The surface wall includes the Ni atom and its two nearest neighbors along the b axis. The two observed impurity states are largely concentrated on Ni and these two neighboring rings, and thus lie almost entirely in the valley or wall. They may represent the 1s and 2s bound states of the Ni impurity, without exchange being a significant factor, or the exchange-split 1s state.

What about the observed oscillations (donor, CuI) and (acceptor, CuII) in the state energies on the length scale b? In the d wave model, these oscillations occur on the length scale of the superconductive gap, that is, the coherence length ~ 6b, and this is the *only* length scale that is present in the model. Thus the observed oscillations on the length scale b flatly contradict the d wave model, and show that a d wave gap is inadequate to describe the data. By far the shortest length scale available is that of domain walls, which indeed is ~ b. Thus the observed oscillations could occur for a wall pseudogap alone, if the Jahn-Teller split gap edge states reversed orbital character between O and Cu(Ni). Another possibility is that both the pseudogap $E_p$ and the superconductive gap $\Delta$ are involved. Because **K** is not a good quantum number near $E_F$, both gaps can be approximated semi-classically as functions of **R,** and both can oscillate near the wall-superconductor interface. The oscillations should be complementary, with $E_p$ being large at Ni and CuII, and $\Delta$ being large near Cu I. Oscillatory behavior involving charge density waves and Jahn-Teller distortions has been identified in CMR manganite layered pseudoperovskites[37].

In the (pseudogap/nanodomain wall) model, the pseudogap of $E_p = 28$ meV is close to the superconductive gap $\Delta$ of 24 meV near optimum doping. Such accidents are distasteful to theory. Is there a reason for this near coincidence, which does not occur in YBCO or LSCO? As we saw in Sec. V and Fig. 4(b), the BSCCO sandwiches can be deformed by the interaction with the STM tip. Now look at the non-planar deformations of the domains and the walls separately. The domains may be flattened by energies associated with the superconductive gap, while the semiconductive domain walls act as hinges. The stiffness of the latter may depend on the magnitude of the pseudogap, which increases as the hinges bend. When the pseudogap in the wall nearly equals the superconductive gap in the domains, the deformation strain energies could be nearly in balance, especially at or near the optimal composition. Thus the near equality of the two gap energies may not be an accident after all.

STM studies of BiO surface inhomogeneities exhibit a high noise level in both the local density of states and the measured gap, which are correlated[38]. The short coherence lengths (14A) indicate, as many workers have previously supposed, a high density of oxygen vacancies in the outer BiO layer. These vacancies give that layer a percolative electronic structure that may mask the underlying self-organized network structure of the superconductive substrate. It is quite interesting that in Fig. 1b, where Fourier (**K**-space) filtering was used to remove topographic variations from the image of a 150x150 A area, the dominant texture is that of (110) filamentary segments. Metallic character is associated with (100) filaments, while (110) texture is usually characteristic of insulating phases[6,7], as one would expect with such a high level of oxygen vacancies.

*Correspondence and request for materials to J. C. Phillips: email, jcphillips@lucent.com

**REFERENCES**

1. D. J. Van Harlingen, Rev. Mod. Phys. **67**, 515 (1995); S. H. Simon and P. A. Lee, Phys. Rev. Lett. **78**, 1548 (1997); Z.X. Shen and D. Dessau, Phys. Rept. **253**, 2 (1995). There are more than 1000 papers that discuss "d-wave" superconductivity; of course, the entire field of HTSC has more than 75,000 papers, and RVB alone has more than 4000 citations.
2. J. C. Phillips and J. Jung, Phil. Mag. B **81**, 745 (2001).

3. J. C. Phillips, Phys. Rev. B **41,** 8968 (1990); Phil. Mag. B, **79**, 527 (1999).

4. M. L. Kulic, Phys. Rep. **338**, 4 (2000).

5. D. Shimada, Y. Shiina, A. Mottate, Y. Ohyagi, and N. Tsuda, Phys. Rev.B **51**, 16495 (1995); D. Shimada, N. Tsuda, U. Paltzer, and F. W. de Wette, Phys. C **298**, 195 (1998). Similar phonon fine structure, that correlates with neutron and Raman phonon spectra, was observed in K-doped $BaBiO_3$ by J. F. Zasadzinski, N. Tralshawala, J. Timpf, D. G. Hinks, B. Dabrowski, A. W. Mitchell, and D. R. Richards, Phys. C **162-164**, 1053 (1989).

6. M. Braden, W. Reichardt, A. S. Ivanov, and A. Y. Rumiantsev, Europhys. Lett. **34**, 531 (1996).

7. J. C. Phillips, Phil. Mag. B **81**, 35 (2001). For a long time it was thought that K-doped $BaBiO_3$ was characterized by [110] tilts of $BiO_6$ octahedra in the insulating phase, while the metallic and superconductive compositions were cubic. M. Braden, W. Reichardt, E. Elkaim, J. P. Lauriat, S. Shiryaev, and S. N. Barilo, Phys. Rev. B **62**, 6708 (2000) have shown by diffraction that the metallic and superconductive compositions are characterized by [100] tilts of $BiO_6$ octahedra, so that the tilt schemes of K- and Pb-doped $BaBiO_3$ are similar. S. Zherlitsyn, B. Luethi, V. Gusakov, B. Wolf, F. Ritter, D. Wichert, S. Barilo, S. Shiryaev, C. Escribe-Fillipini, and J. L. Tholence, Eur. J. Phys. B **16**, 59 (2000) with ultrasonic measurements found softening of both $c_{11}$ and $c_{44}$ in two superconductive compositions of K-doped $BaBiO_3$ between 200K and 50K, which is correlated with the [100] tilts, but which we ascribe to hysteretic (100) ordering of K. This is another example of the intimate relationship of [100] or (100) dopant ordering and HTSC in perovskite and pseudo-perovskite phases.

8. D. Haskel, E. A. Stern, D. G. Hinks, A. W. Mitchell, and J. D. Jorgensen, Phys. Rev. B **56**, R520 (1997).

9. W. K. Neils and D. J. Van Harlingen, Physica B **284-288**, 587 (2000).

10. E.W. Hudson, S.H. Pan, A.K. Gupta, K.W. Ng and J.C. Davis, Science **285**, 88 (1999).

11. S.H. Pan, E.W. Hudson, K. M. Lang, H. Elsaki, S. Uchida, and J.C. Davis, Nature **403**, 746 (2000).

12. S.H. Pan *et al.,* Bull. Am. Phys. Soc. **46** (1), 1043 (2001).

13. J. Etheridge, Phil. Mag *A* **73**, 643 (1996).

14. H. Darhmaoui and J. Jung, Phys. Rev. B **53**, 14620 (1996); Phys. Rev. B **57**, 8009 (1998).

15. J. C. Phillips and J. Jung, LANL Cond-Mat/0103167 (2001).

16. C. Howard, P. Fournier and A. Kapitulnik, LANL Cond-Mat/0101251(2001).

17. M. Masuda and T. Nishinaga, J. Cryst. Growth **198/199**, 1098 (1999).

18. N. Miyakawa, J. F. Zasadzinski, L. Ozyuzer, P. Guptasarma, C. Kendziora, D. G. Hinks, T. Kaneko, and K. E. Gray, Physica C **341-348**, 835 (2000).

19. M. Oda, T. Matsuzaki, N. Monono, and M. Ido, Physica C **341-348**, 847 (2000).

20. A. F. Hebard and A. T. Fiory, Phys. Rev. Lett. **44**, 291 (1980).

21. D. G. Wiesler, M. Suzuki, P. C. Chow, and H. Zabel, Phys. Rev. B **34**, 7951 (1986); D. G. Wiesler, H. Zabel, and S. M. Shapiro, Zeit. Physik B **93**, 277 (1994).

22. J. P. Rodriguez, Phys. Rev. B **62**, 9117 (2000).

23. T. Oki, N. Tsuda, and D. Shimada, Physica C **341-348**, 913 (2000).

24. C. C. Homes, T. Timusk, D. A. Bonn, R. Liang, and W. N. Hardy, Can. J. Phys. **73**, 663 (1995).

25. W. Reichardt, N. Pyka, L. Pintschovius, B. Hennion and G. Collin, Phys. C **162-164**, 464 (1989).

26. G. Blumberg, M. Kang, M. V. Klein, K. Kadowaki, and C. Kendziora, Science **278**, 1427 (1997).

27. H. L. Liu, G. Blumberg, M. V. Klein, P. Guptasarma, and D. G. Hinks, Phys. C **341-348**, 2181 (2000).

28. J. C. Phillips, Phil. Mag. B **79**, 527 (1999); in press (2001).

29. M. F. Thorpe, M. V. Chubinsky, D. J. Jacobs and J. C. Phillips, J. Non-Cryst. Solids, **266-269**, 859 (2000).

30. E. Kaldis, J. Röhler, E. Liarokapis, N. Poulakis, K. Conder, and P.W. Loeffen, Phys. Rev. Lett. **79**, 4894 (1997).

31. D. G. Hinks, J. D. Jorgensen, J. L. Wagner, O. Chaimssem, and B. Dabrowski, Bull. Am. Phys. Soc. **46** (1), 1329 (2001).

32. M. A. Topinka, B. J. Leroy, R. M. Westerfelt, S. E. J. Shaw, R. Fleischmann, E. J. Heller, K. D. Maranowski, and A. C. Gossard, Nature **410**, 183 (2001).

33. D. Orgad, S. A. Kivelson, E. W. Carlson, V. J. Emery, X. J. Zhou, and Z.X. Shen, Phys. Rev. Lett. **86**, 4362 (2001).

34. E.W. Hudson, K. M. Lang, V. Madhavan, S.H. Pan, H. Eisaki, S. Uchida, and J.C. Davis, Nature **411**, 920 (2001).

35. M. I. Salkola, A. V. Balatsky, and J. R. Schrieffer, Phys. Rev. B **55**, 12648 (1997).

36. Like all the figures of ref. 34, this is best viewed at LANL cond/mat 0104237.

37. Y.-D. Chuang, A. D. Gromko, D. S. Dessau, T. Kimura, and Y. Tokura, Science **292**, 1509 (2001). In these materials the metal-insulator transition appears to be simply percolative, without an intermediate phase. Some self-organization may occur, but it is incomplete, and the first and second transitions are not separated. Presumably this reflects the low dopant (Sr) diffusivity, together with low oxygen diffusivity.

38. S.H. Pan, J. P. O'Neal, R. L. Bladzey, C. Chamon, H. Ding, J. R. Engelbrecht, Z. Wang, H. Eisaki, S. Uchida, A. K. Gupta, K.-W. Ng, E.W. Hudson, K. M. Lang, and J.C. Davis, LANL cond/mat 0107347.

**Figure Captions**

Note for arXiv file. Figs. 1-3 are included only for the reader's convenience. To save space for this file they are omitted here. The original figures can be found as follows: Fig. 1, ref. 3; Fig.2, ref.5; Fig. 3, ref. 6.

Fig. 1. The basic idea of the filamentary paths in the quantum percolative model[3] for YBCO. The positions of the **I**nsulating **N**anodomain **W**alls in the $CuO_2$ layers are indicated, together with the **R**esonating **T**unneling **C**enters in the semiconductive layer, and oxygen vacancies in the $CuO_{1-x}$ chains. Giant e-p interactions are associated with the **R.T.C.**, where the interactions with LO c-axis phonons are especially large. The **I.N.W.** are perovskite-specific. The *sharp bends* in the filamentary paths are responsible for the broken symmetry that admixes ab planar background currents with c-axis LO phonons.

Fig. 2. The electron-phonon spectral function $\alpha^2 F$ measured in the best BSCCO tunnel junction, compared with the phonon spectrum measured by neutron scattering and the calculated quasiparticle density of states[5]. Sharp edge peaks above the frame are shown as inset. These data show that the energy gap is isotropic (s wave), and that electron-phonon interactions cause HTSC. In this sample the doping is optimal.

Fig. 3. LO dispersion curves[6] of superconductive $Ba_{0.6}K_{0.4}BiO_3$ (filled circles) and $BaKBiO_3$ (empty circles). Note that all the curves are quite flat, except for the kink in the HTSC [100] K-doped mode near [0.2,0,0]. Because this anomaly is absent in the undoped sample, it seems obvious that this anomaly is associated with partial ordering of the dopants in (100) planes[7]. This partial ordering gives rise to anisotropic screening of LO phonon interactions, with singular behavior reflecting the wave length of the partial ordering.

Fig. 4. Each BSCCO layer is represented by a line, and the letters represent the spacings of the molecular layers bound by Van der Waals forces. Because of interactions with the contacts, these spacings satisfy the following inequalities: d < c < e < f.

Fig. 5. Sketch of doping dependence of $\Delta(0)$ and $T_c$ as measured with break and point contact junctions[18,19]. Optimal doping occurs near p = 0.17. The mystery is why $\Delta(0)$ grows as p decreases in the underdoped region below 0.17, instead of decreasing, as $T_c$ does. In the text this is explained as the result of interlayer decoupling. The upper abcissa is a guess at the interlayer coupling strength $\sigma$, a parameter similar to $J_\perp/J_{||}$ in xy spin models[22]. The circle marks the centered optimal composition and energy gap of the best BSCCO epoxy tunnel junction[5] shown in Fig. 2. The square marks the energy gap observed in a recent STM experiment[16]. This sample gives the largest spatially inhomogeneous effects, and it lies at the low-density edge of the underdoped superconductive region.

Fig. 6. Sketch of Δ(T) in underdoped samples, showing collapse[21] near $T_c$, and in optimally and overdoped samples, showing almost mean-field (BCS) behavior[18].

Fig. 7. A compact summary of the differences between continuum (**k**-space) and filamentary (**R**-space) models of normal and superconductive metals.

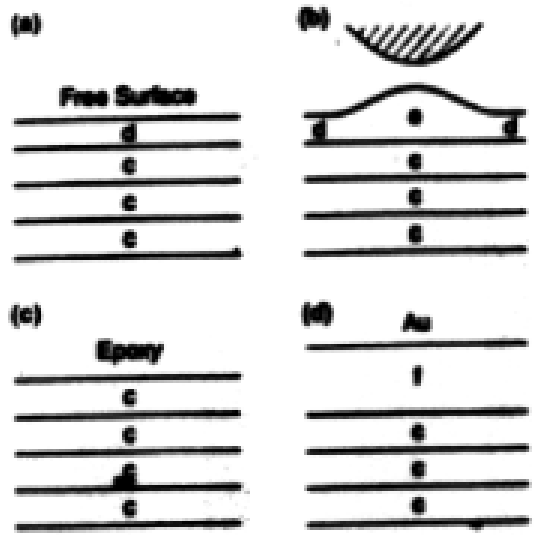

Fig. 4

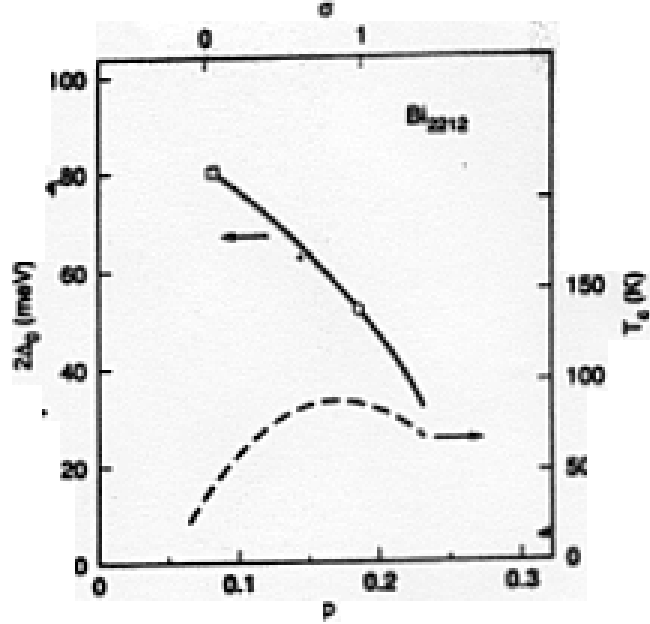

Fig. 5

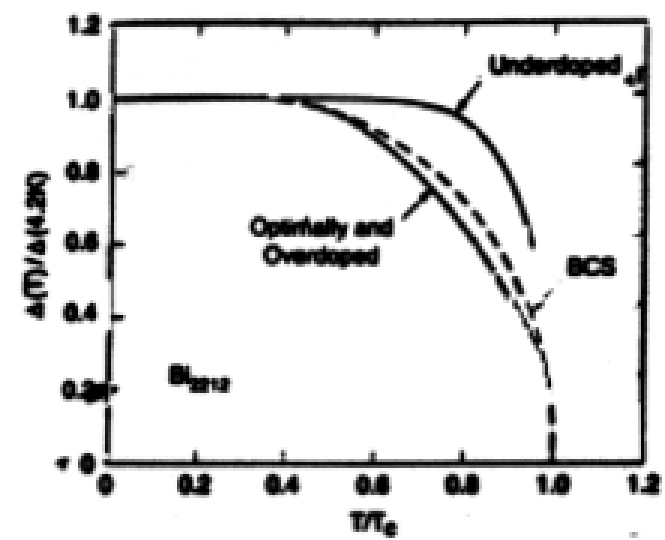

Fig. 6

Fig. 7